\documentclass[%
superscriptaddress,
 amsmath,amssymb,
]{article}

\usepackage[margin=1in]{geometry} 
\usepackage{graphicx}
\usepackage{dcolumn}
\usepackage{bm}

\usepackage{braket}

\usepackage{siunitx,upgreek}

\usepackage{amsmath}
\usepackage{amssymb}
\usepackage{amsthm}

\usepackage{hyperref}

\usepackage{comment}
\usepackage{todonotes}
\usepackage{amsmath}
\usepackage{tikz}
\usetikzlibrary{arrows, positioning, fit, calc}

\usepackage{listings}
\usepackage{xcolor}

\title{Secure Multi-Party Biometric Verification using QKD assisted Quantum Oblivious Transfer}
\author{Mariana F. Ramos$^1$\thanks{mariana.ferreira-ramos@ait.ac.at} \and Michael Hentschel$^1$ \and Federico Valbusa$^1$ \and Costin Luchian$^1$ \and Martin Achleitner$^1$ \and Alessandro Trenti$^1$ \and Marie-Christine Slater$^1$ \and Mariano Lemus$^2$ \and Thomas Lorünser$^1$ \and Hannes Hübel$^1$}

\date{
	$^1$ AIT Austrian Institute of Technology, Center for Digital Safety \& Security / Security \& Communication Technologies, 1210 Vienna, Austria. \\ %
	$^2$ Instituto de Telecomunica\c{c}\~oes, Optical Quantum Communications Group, 3810 - 193 Aveiro, Portugal\\[2ex]%
	\today
}
\begin{document}
\maketitle

\begin{abstract}
We present a practical implementation of a secure multiparty computation application enabled by quantum oblivious transfer (QOT) on an entanglement-based physical layer. The QOT protocol uses polarization-encoded entangled states to share oblivious keys between two parties with quantum key distribution (QKD) providing authentication. Our system integrates the post-processing for QKD and QOT, both sharing a single physical layer, ensuring efficient key generation and authentication. Authentication involves hashing messages into a crypto-context, verifying tags, and replenishing keys through a parallel QKD pipeline, which handles both key post-processing and authentication. Oblivious keys are generated over 12.9 km with a channel loss of 8.47 dB. In a back-to-back setup, a QOT rate of $9.3\times10^{-3}$ OTs/second is achieved, corresponding to 1 minute and 48 seconds per OT, primarily limited by the entanglement source. Using pre-distributed keys improved the rate to 0.11 OTs/second, or 9.1 seconds per OT. The considered QOT protocol is statistically correct, computationally secure for an honest receiver, and statistically secure for an honest sender, assuming a computationally hiding, statistically binding commitment, and a verifiable error-correcting scheme. A practical use case is demonstrated for privacy-preserving fingerprint matching against no-fly lists from Interpol and the United Nations. The fingerprint is secret-shared across two sites, ensuring security, while the matching is performed using the MASCOT protocol, supported by QOT. The application required 128 OTs, with the highest security achieved in 20 minutes and 39 seconds. This work demonstrates the feasibility of QOT in secure quantum communication applications.
\end{abstract}

\maketitle


\section{Introduction}
In an era where global travel security and privacy concerns are at the forefront, ensuring robust and secure identity verification systems is crucial. Biometric-based systems have become increasingly popular to ensure border safety at airports and other security-critical infrastructures, where entities like national governments, the United Nations, and INTERPOL must collaborate to verify identities and prevent unauthorized access. However, these systems face significant challenges in safeguarding sensitive biometric data, particularly when multiple parties are involved in cross-jurisdictional biometric verification. The need to verify identity while maintaining individuals' privacy introduces complexities in data security and trust management. 

Multi-party computation (MPC) is well-suited for this scenario as it meets both key objectives: it allows confidential pooling of No-Fly lists among organizations that prefer not to openly share data, and it ensures the privacy of regular travelers by safeguarding their biometric information \cite{treiber_data_2022}. Only in the case of a match will relevant data be revealed for further investigation, while the fingerprints of non-matching travelers remain fully protected from disclosure. Oblivious transfer (OT) is a basic primitive for secure MPC when trust cannot be guaranteed among the parties. OT implementations aim to secure information transfer between parties while ensuring that none of them gains more information than they are entitled to. 

Classical OT protocols rely on computational assumptions associated with public-key cryptography. It is well known that the advance of quantum computers poses a threat to the public-key primitives traditionally used to construct OT (e.g., RSA)\cite{santos_quantum_2021} and, although several alternatives based on post-quantum cryptography have been proposed \cite{branco2021, mi2018, mansy2019}, all of them require the existence of \textit{trapdoor} one-way functions. Such functions are defined over mathematically rich structures and their properties are less understood as compared to their private-key cryptography counterparts. 

Quantum OT implemented within the noisy storage model provides security benefits over classical OT by assuming limited quantum memory capabilities for an adversary. This approach offers future-proof security and post-quantum resilience, ensuring robustness against state-of-the-art quantum adversaries \cite{erven_experimental_2014}. However, these benefits come with performance trade-offs, including longer execution times, lower key generation rates, and stringent error rate requirements. As quantum memories continue to improve, the security assumptions of the noisy storage model may be compromised and require substantial updates by sacrificing even more the performance \cite{erven_experimental_2014, ng_experimental_2012}. In this way, it is to be expected that with technology development, the performance of quantum OT in the noisy storage model decreases to keep the same level of security. An alternative is the hybrid implementation of the quantum OT proposed in \cite{lemus_generation_2020}. The security of hybrid quantum OT relies on the principles of quantum physics for key exchange between parties, combined with a computationally hiding, statistically binding commitment classical bit commitment phase. This ensures the honesty of both the receiver and the transmitter, thereby guaranteeing privacy for both parties \cite{halevi_practical_1998}. Although currently achieving a lower performance performance than the quantum OT in the noisy storage model, it is expected that with technology development the performance increases due to an increase in computational power, contrary to the previous protocol. Additionally, the commitment only relies on the cryptography strength of one-way functions, which is considered more secure than public-key cryptography.

In this paper, we present a full-stack quantum-enabled biometric verification system that integrates entanglement-based quantum OT and Quantum Key Distribution (QKD) to enhance the security of multi-party biometric authentication processes. Specifically, we focus on a use case where organizations such as the United Nations and INTERPOL collaborate to match the biometric data of passengers at airports on No-Fly databases \cite{strobl2022beimetricdatasecurity,mader_towards_2024, groote_formal_2022}. Our system allows these organizations to securely compute whether a fingerprint presented at an airport matches an entry in a shared biometric database, without revealing the identity of the passenger to any party involved in the computation. Through this work, we aim to bridge the gap between quantum cryptographic theory and practical application. To this end, we implement a complete quantum communication stack, from the physical layer to the application layer, leveraging entanglement-based quantum systems for secure key distribution and oblivious transfer. 

This paper is organized into five sections. Section II outlines the quantum OT protocol. Section III introduces the fully integrated software stack, which combines QKD-based authentication with QOT and its integration into an MPC application. Section IV details the methods and presents the main results. Finally, Section V summarizes the key conclusions of this work.

\section{Protocol Description}
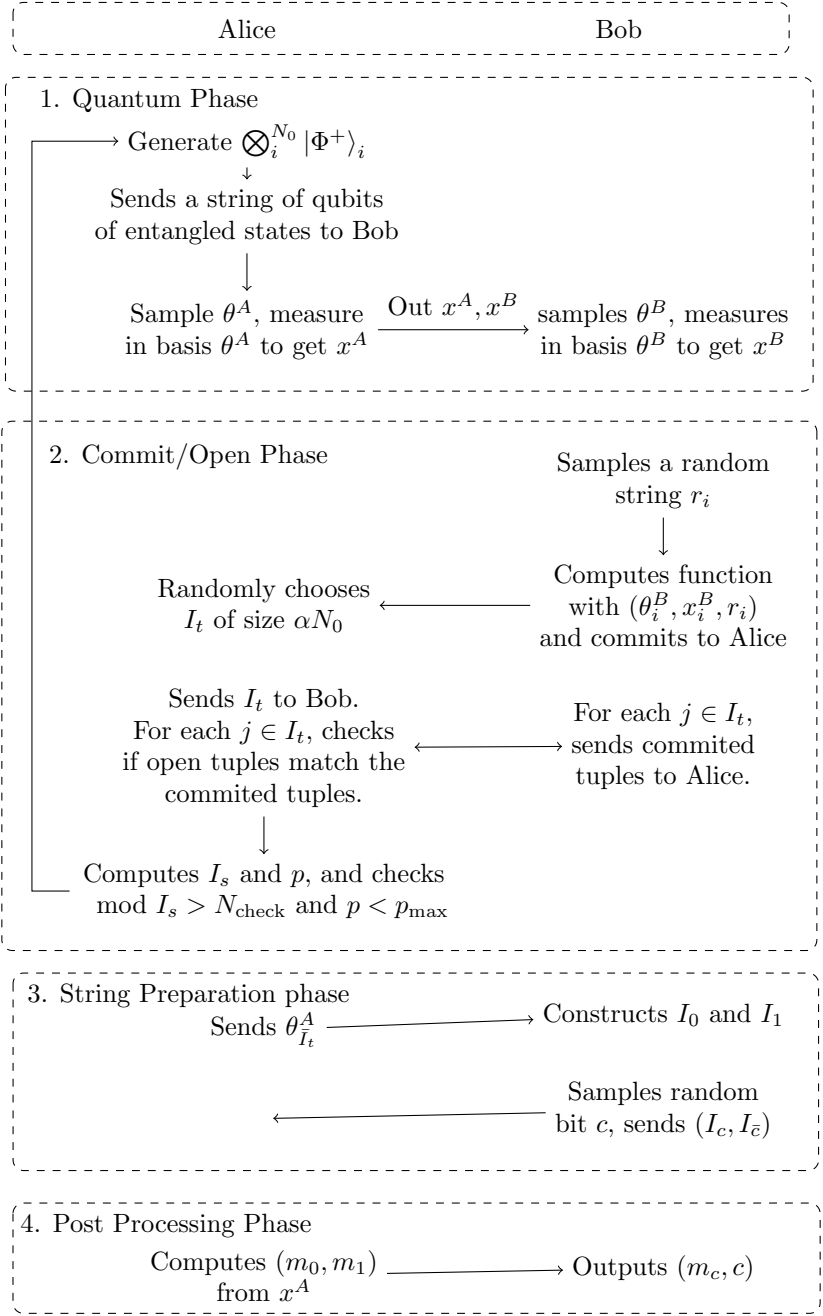
\begin{figure}[t!] 
\centering
\begin{tikzpicture}[node distance=1cm, auto, scale=1, transform shape] 

    \node (alice) {Alice};
    \node[right=4cm of alice] (bob) {Bob};

    \node[below of=alice, node distance=1.5cm] (genstates) {Generate $\bigotimes_{i}^{N_0} \ket{\Phi^+}_i$};
    \node[below of=genstates,align = center] (sendbob) {Sends a string of qubits \\ of entangled states to Bob};
    
    \draw[->] (genstates) -- (sendbob);
    
    \node[below=0.5cm of sendbob, align=center] (measA) {Sample $\theta^A$, measure \\ in basis $\theta^A$ to get $x^A$};
    \node[right=2cm of measA, align = center] (measB) {samples $\theta^B$, measures \\ in basis $\theta^B$ to get $x^B$};
    \draw[->] (measA) -- (measB) node[midway, above] {Out $x^A, x^B$};
    \draw[->] (sendbob) -- (measA);

    \node[below=1cm of measB, align=center] (randomstr) {Samples a random \\ string $r_i$};
    \node[below=0.5cm of randomstr,align=center] (commitB) {Computes function \\with $(\theta^B_i, x^B_i, r_i)$ \\ and commits to Alice};
    \draw[->] (randomstr) -- (commitB);

    \node[left=2cm of commitB, align=center] (subset) {Randomly chooses \\ $I_t$ of size $\alpha N_0$};
    \node[below=0.5cm of subset, align=center] (checkA) {Sends $I_t$ to Bob. \\ For each $j \in I_t$, checks\\ if open tuples match the \\ commited tuples. };
    \node[below=0.5cm of commitB,align=center] (checkB){For each $j\in I_t$, \\sends commited \\tuples to Alice.};
    \draw[->] (commitB) -- (subset);
    \draw[<->] (checkA) -- (checkB);

    \node[below=0.5cm of checkA, align=center] (compA) {Computes $I_s$ and $p$, and checks \\ $\mod{I_s}> N_{\textnormal{check}}$ and $p<p_{\textnormal{max}}$};
    \draw[->] (checkA) -- (compA);
    \draw[->] (compA.west) -- ++(-0.5cm, 0) |- (genstates);

    \node[below=1cm of compA] (sep1) {Sends $\theta^A_{\Bar{I}_t}$};
    \node[below=2.6cm of checkB] (constructI0) {Constructs $I_0$ and $I_1$};
    \draw[->] (sep1) -- (constructI0);

    \node[below=0.5cm of constructI0, align=center] (randc) {Samples random \\bit $c$, sends $(I_c, I_{\Bar{c}})$};
    \node[below=0.7cm of sep1, align=center] (randcA) {       };
    \draw[->] (randc) -- (randcA);

    \node[below=1.5cm of randcA, align=center] (m0m1) {Computes $(m_0, m_1)$ \\ from $x^A$};
    \node[below=1.3cm of randc] (mc) {Outputs $(m_c, c)$};
    \draw[->] (m0m1) -- (mc);

    \draw[rounded corners, dashed] ($(genstates.north west) + (-1.5,0.5)$) rectangle ($(measB.south east) + (0.2,-0.3)$);
    \draw[rounded corners, dashed] ($(randomstr.north east) + (0.5,0.3)$) rectangle ($(compA.south west) + (-0.9,-0.3)$);
    \draw[rounded corners, dashed] ($(sep1.north west) + (-2.5,0.4)$) rectangle ($(randc.south east) + (0.5,-0.3)$);
    \draw[rounded corners, dashed] ($(m0m1.north west) + (-1.7,0.5)$) rectangle ($(mc.south east) + (0.75,-0.3)$);
    \draw[rounded corners, dashed] ($(alice.north west) + (-2.6,0.1)$) rectangle ($(bob.south east) + (1.83,-0.1)$);

    \node at ($(genstates.north) + (-1.3,0.2)$) {1. Quantum Phase};
    \node at ($(subset.north) + (-1,1.5)$) {2. Commit/Open Phase};
    \node at ($(sep1.north) + (-1,0.1)$) {3. String Preparation phase};
    \node at ($(m0m1.north) + (-1.3,0.2)$) {4. Post Processing Phase};
    
\end{tikzpicture}
\caption{Protocol steps for Randomized Oblivious Transfer (ROT)}
\label{fig:rot_protocol}
\end{figure}

We consider the $n$-bit Randomized Oblivious Transfer (ROT) probabilistic protocol from Lemus et al. \cite{lemus_generation_2020}. This protocol is established between two parties, a sender Alice and a receiver Bob, and returns two uniformly distributed strings $m_0$, $m_1$ to the sender, and a random bit $c$ together with message $m_c$ to the receiver. The protocol is considered secure against a dishonest sender if they are unable to gain any information about the value of $c$. Similarly, the protocol is secure against a dishonest receiver if, by the end of the protocol, they remain unaware of at least one of the two strings $m_0$ and $m_1$.
The protocol is split into four phases:
\begin{enumerate}
    \item \textit{Quantum Phase}: Alice generates a sequence of entangled states $\bigotimes_{i}^{N_0} \ket{\Phi^+}_i$ and sends one qubit to Bob. Then she samples the string $\theta^{A}\in \{0,1\}^{N_0}$ ($N_0$ is a protocol parameter that indicates the number of shared entangled states) and for each $i=1,\ldots,N_0$ performs a measurement in the basis $\theta^{A}_i$ on her qubit of $\ket{\Phi^+}_i$ to obtain the outcome string $x^{A}$. Bob samples the string $\theta^{B}\in \{0,1\}^{N_0}$ and for each $i \in I = \lbrace 1,\ldots,N_0 \rbrace$ performs a measurement in the basis $\theta^{B}_i$ on his qubit of $\ket{\Phi^+}_i$ to obtain the outcome string $x^{B}$.
    \item \textit{Commit/Open phase}:  For each $i \in I$, Bob samples a random string $r_i$, computes the commitment function with the inputs $(\theta^{B}_i,x^{B}_i,r_i)$ and commits to Alice, as described in Appendix \ref{ourBSC}. Alice randomly chooses a test subset $I_{t}$ of size $\alpha N_0$ and sends $I_{t}$ to Bob, where $\alpha$ is the parameter estimation sample ratio. For each $j \in I_{t}$, Bob sends the corresponding committed tuples to Alice. For each $j \in I_{t}$, Alice checks that the opened tuples match the committed tuples. Then, Alice computes the set $I_{s}=\lbrace j \in I_{t}  \vert \theta^{A}_j = \theta^{B}_j \rbrace$ and the quantity
      \begin{equation}
          p=\frac{1}{|I_{s}|}\sum_{j \in I_{s}}(x^{A}_j \oplus x^{B}_j),
      \end{equation}
      and checks that $|I_{s}| \geq N_{check}$ and $p \leq p_{\text{max}}$, where $N_{check}$ refers to the minimum check set size, and $p_{max}$ is the maximum allowed qubit error rate. If any of the checks fail, Alice aborts the protocol \cite{lemus_performance_2025}.
    \item \textit{String separation phase}: In this stage, Bob and Alice perform sifting. Alice sends $\theta^{A}_{\Bar{I}_t}$ to Bob. Bob constructs the set $I_0$ by randomly selecting $N_{\text{raw}}$ indices $i \in \Bar{I}_t$ for which $\theta^{A}_i = \theta^{B}_i$. Similarly, he constructs $I_1$ by randomly selecting $N_{\text{raw}}$ indices $i \in \Bar{I}_t$ for which $\theta^{A}_i \neq \theta^{B}_i$. The raw string block size is obtained from $N_{\text{raw}} = (\frac{1}{2}- \delta_2) (1 - \alpha) N_{0}$, where $\alpha$ is the parameter estimation ratio and $\delta_2$ is a statistical tolerance parameter. He then samples a random bit $c$, and sends the ordered pair $(I_c, I_{\Bar{c}})$ to Alice, where $c$ corresponds to his bit choice. If Bob is not able to construct $I_0$ or $I_1$, he aborts the protocol.
    \item \textit{Post processing phase}: In this phase, Bob and Alice perform error correction and privacy amplification. Alice outputs $(m_0,m_1)$, where $m_0=\bigoplus_{j \in I_c}x_{j}^{A}$, and $m_1=\bigoplus_{j \in I_{\Bar{c}}}x_{j}^{A}$. Bob outputs $(m_c,c)$, where $m_c=\bigoplus_{j \in I_0}x_j^{B}$.
\end{enumerate}
In the asymptotic limit of large $N_0$, the ROT secret key rate can be written as follows:
 \begin{equation}
    R_{\textnormal{key}}^{\textnormal{asympt}} = (1 - \alpha) \left(\frac{1}{2}- \delta_2\right) \left( \frac{1}{2} - \delta_2 - h\left( \frac{p_{\textnormal{max}} + \delta_1}{\frac{1}{2} - \delta_2}\right) - f \cdot h(p_{\textnormal{max}}+\delta_1) \right),
\label{eq:keyrate}
\end{equation}   
where $0 < \alpha < 1$ is the parameter estimation sample ratio, $\delta_1, \delta_2 $ are statistical tolerance parameters, $0 < p_{\text{max}} \leq 1/2$ is the maximum qubit error rate, $h(\cdot)$ denotes the binary entropy, and $f$ is the error correction efficiency.

\begin{table}[h!]
\centering
\begin{tabular}{@{}ccc@{}}
\hline
\textbf{Parameter}                & \textbf{Symbol}      & \textbf{Value}        \\ \hline
Sample ratio                  & $\alpha$                  & 0.35                \\ 
Efficiency of IR                  & $f$                  & 1.0270                \\ 
Commitment biding security        & $\varepsilon_{bind}$ & $2^{-253}$            \\ 
Number of shared entangled states & $N_0$                & $3.2 \times 10^6$    \\ 
Security level                    & $\varepsilon_{tot}$  & $2.35 \times 10^{-8}$ \\ 
Maximum QBER                      & $p_{max}$            & $1.4 \%$              \\ 
Statistical parameter 1           & $\delta_1$           & $1.34 \times 10^{-2}$ \\ 
Statistical parameter 2           & $\delta_2$           & $5 \times 10^{-3}$    \\ 
IT verifiability security         & $\varepsilon_{IR}$   & $2^{-96}$             \\ \hline
\end{tabular}
\caption{Protocol parameters}
\label{tab:param}
\end{table}

Table \ref{tab:param} summarizes the key parameters utilized for implementing the protocol.

\section{A Fully Integrated Software Stack Combining QKD and QOT} 
\begin{figure*}[t]
\centering
\includegraphics[scale=0.8]{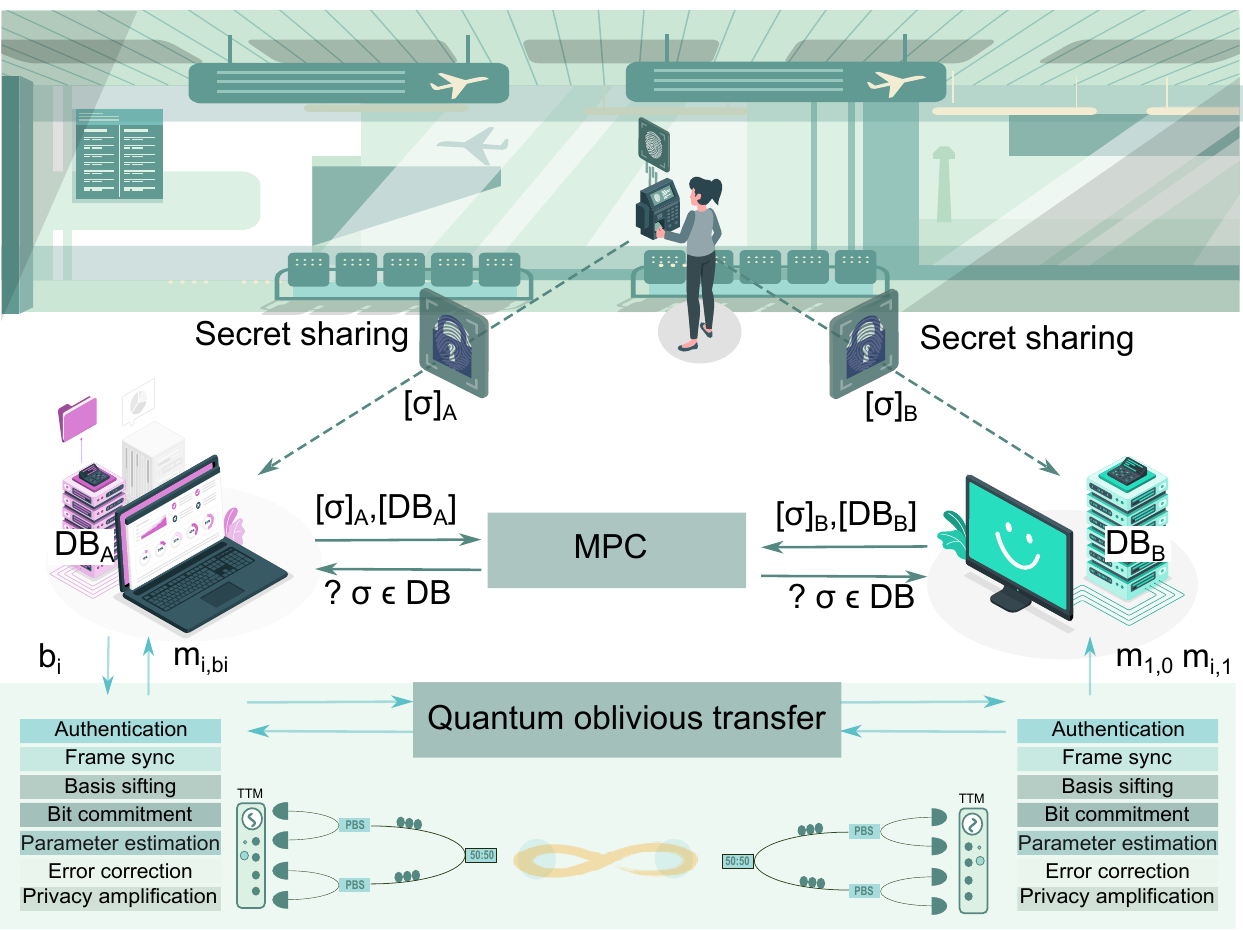}
\caption{Full implementation of a secure multiparty computation application for biometric data matching, supported by a quantum OT protocol based on entanglement distribution and secret sharing. The system enables two distant entities to securely compare biometric inputs with distributed databases while preserving the privacy of both the input data and the contents of each database. }
\label{fig:smc}
\end{figure*}
In this section, we present the complete software stack implemented to integrate QKD and QOT.
We consider a scenario in which two distant parties aim to jointly compute a function on a shared database while preserving the privacy of their respective inputs and the database itself. The parties neither trust each other nor the communication channel. For this scenario, an MPC application is implemented, supported by a QOT protocol that leverages QKD-based authentication.


\subsection{Post-processing}

The overall setup of the QKD and QOT post-processing pipeline is shown in Fig. \ref{PP_SETUP}. It comprises several modules sequentially connected to implement the complete protocol. A separate instance of the pipeline operates at either site: the sender (Alice) and the receiver (Bob). Most of the modules require classical communication between the two sites over an authenticated channel. This channel is assumed to be error-free but otherwise fully public (e.g. an Ethernet connection). Authentication is handled by the post-processing software, which uses a small pre-shared secret and periodic quantum key replenishment provided by the QKD lane of the pipeline.

The application front-end is facilitated by a client object that serves two purposes: requesting a number of OTs of a given size (green connection) and retrieving them from the pipeline's output (orange connection). On the receiver side, each OT request is accompanied by a bit-choice for each OT, specifying which OT-half is the correlated one. On the sender side, only the correlated OT-half is obtained. The upper sequence of modules represents the protocol steps for quantum OT, while the lower sequence handles the processing of QKD keys to replenish the authentication keys.

\subsubsection{Raw Data Buffering}
\begin{figure*}[t]
\centering
\includegraphics[scale=0.75]{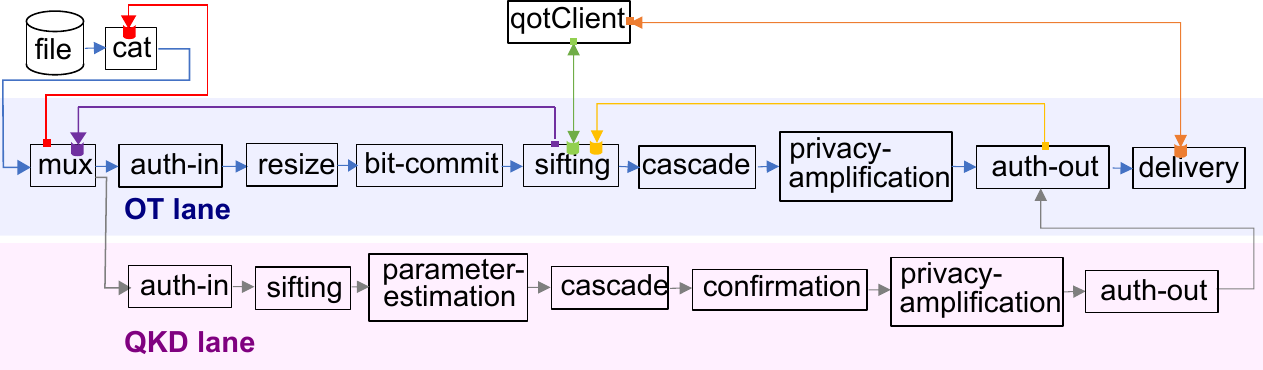}
\caption{The post-processing pipeline running at both parties.}
\label{PP_SETUP}
\end{figure*}

In order to assess the post-processing performance, we pre-recorded raw data from the physical setup and pre-sifted the data for coincidences. This dataset is stored in an appropriate input format (\textit{file} in Fig. \ref{PP_SETUP}), ready for use by the post-processing pipeline. When raw keys are requested (red connection), the \emph{cat} module retrieves them from the file storage and forwards them to the \emph{mux} module. This module is triggered by the \emph{ot-sifting} module (purple connection), which serves as the central hub for key requests. The \emph{mux} module then  directs the keys either to the OT-lane or the QKD-lane, depending on the specific request.

\subsubsection{Authentication}
To authenticate the entire message flow between the peer modules, each key is assigned a handle by the \emph{auth-in} module at the start of the pipeline. All messages exchanged between peer modules are hashed into this context. At the end of the pipeline, the \emph{auth-out} module hashes the pre-shared secret onto the message digest to generate the final authentication tag. These tags are compared between Alice and Bob to verify the authenticity of the classical communication. Once the pre-shared secret is depleted, it must be replenished with additional key material. Keys generated in the OT-pipeline cannot be used for this purpose because only one half is correlated and Alice does not know which. Instead, the parallel QKD-pipeline supplies the required key material. When the \emph{OT-auth-out} module requires additional key material, it sends a request to \emph{OT-sifting} module (yellow connection), which forwards the request to the \emph{mux} module. The \emph{mux} module then redirects raw data to the QKD-pipeline, where standard QKD post-processing is performed using the same raw key source as the OT protocol. Authentication within the QKD-pipeline is carried out similarly, but the pipeline is self-sufficient, using processed keys for authentication replenishment. The final processed keys are then forwarded to the \emph{ot-auth-out} module, allowing it to continue operation.

\subsubsection{Bit Commitment and Parameter-estimation}
An essential requirement of the 1-2 OT protocol is that Bob must be able to have access to only one of the two messages. If he possesses a quantum memory, he could store his qubits and measure them only after Alice discloses her measurement bases, thereby gaining both messages. To ensure honesty, Bob must commit to all his measurement choices before Alice announces her bases. A fraction of these commitments are then opened for verification by Alice. To achieve this, we implement a variant of Naor's bit commitment scheme (see \ref{ourBSC}). This scheme requires no additional assumptions to prove the unconditional binding property, which is essential for the QOT security proof \cite{lemus_generation_2020}, and enhances communication efficiency compared to Naor's original protocol \cite{RLV24}.
A detailed description of the protocol is provided in the Appendix \ref{sec:bitcommitment}.

The measurements disclosed by Bob during the bit commitment can also be used for statistical analysis to assess security parameters, including the epsilons for correctness, hiding and binding (see Appendix \ref{sec:security}), as well as the maximum secure key length, see (Eq. \ref{eq:keyrate}). A key factor in evaluating the security of the OT protocol is the size of the raw input key blocks. Achieving a higher security parameter ($\varepsilon_{tot}$) requires larger raw input key block sizes. To allow flexibility in adjusting this size, a \emph{resize} module is included before the main protocol begins. This module accumulates and/or splits keys as needed to meet the required block size.

\subsubsection{Sifting}
A key aspect of the protocol is that Bob's measurement bases align with Alice's approximately half of the time. This results in a correlated key block and an uncorrelated one. The purpose of the \emph{OT-sifting} module is to separate these blocks based on Alice's announcement of her measurement bases and arrange the blocks according to Bob's (unknown to Alice) bit-choice. Since this step introduces the bit-choice, it is the point where the client connects to issue OT requests. The module then triggers the release of key material at the pipeline's input. Additionally, when the authentication module requires new key material, the \emph{OT-sifting} module also initiates a QKD key release.

\subsubsection{Error Correction}

To correct errors in the correlated key, we use the Cascade algorithm. This choice over LDPC codes is based on two main considerations: for the low QBER values required by the OT protocol ($\sim 1 \%$) Cascade achieves a very high efficiency ($> 99 \%$), and Cascade allows for operation on variable-length block sizes, which is challenging  with LDPC codes. 
Because Cascade is highly interactive and Alice must not learn which of the two keys is the correlated one, certain modifications to the algorithm are necessary. During the protocol execution, Alice and Bob repeatedly exchange parity check sums for different sub-blocks to identify bit errors. In our modified Cascade, Alice computes and sends parities for both keys. Bob calculates the parity differences for the correlated key only, applies them to his key, and sends the results back to Alice. Alice does not change her keys but uses Bob's parity checks to select matching sub-blocks for the next steps of the protocol. This ensures that the correlated key is corrected without revealing to Alice which key is the correlated one.

\subsubsection{Privacy-amplification}
During parameter-estimation, in the bit commitment step, we calculate the maximum secure key length based on the protocol parameters and the actual error correction efficiency. As long as this value exceeds the requested OT length, we can proceed by reducing the correlated key to the final OT length using Toeplitz-hashing.

\subsection{Integration of QOT with MPC applications}

It has been demonstrated that OT is universal, in the sense it can be used to implement any cryptographic task \cite{Kilian88}.
Moreover, OT enables advanced cryptographic functionalities, such as secure computation on encrypted data through secure multiparty computation.
We investigate methods for integrating QOT with existing MPC protocols to efficiently and practically harness quantum advantages.
It is important to note that a QOT works essentially as a random OT (ROT), analogous to how QKD generates random keys but cannot directly transmit a given key.
Efficient protocols exist to convert any ROT into standard OT with perfect security, and vice versa \cite{Wolf2006}. For instance, in a straightforward approach, the sender can use the ROT output to encrypt two messages, which are then transmitted to the receiver.

MPC is defined as an ideal functionality in which parties provide inputs to a trusted entity that computes a function on the input data and returns the outputs to the respective parties. Over the last few decades, several protocols have been developed to implement MPC ~\cite{Yao1986,GMW87,spdz}.
The key security properties of MPC are correctness and input privacy. Correctness ensures that the function is computed accurately, while input privacy guarantees the confidentiality of the parties' data.
Depending on the protocol, the security parameters can provide protection against different types of adversaries.

We have identified three potential approaches to leverage the benefits of QOT into existing MPC protocols, as summarized in Table \ref{tab:mpc_ot_integration}.
Yao's based protocols, which use garbled circuits as their core concept, rely on OT to transmit inputs from the garbler to the evaluator ~\cite{Yao1986}.
The security of the garbled circuits depends on the hardness of symmetric ciphers. By incorporating our hybrid QOT, we can eliminate reliance on structured hardness assumptions across the entire protocol.
In GMW protocols, the inputs are secret-shared and OT is employed to evaluate non-linear operations in the function ~\cite{GMW87}.
Since secret sharing provides Information-Theoretic Security (ITS), replacing classical OT with the QOT enhances security by eliminating dependence on asymmetric cryptography, which is the only non-ITS component.
SPDZ, an extension of the BGW method, achieves security against dishonest majorities ~\cite{BGW88}.
It uses secret sharing and operates with ITS in its online phase but requires preprocessing to generate randomized multiplication triples.
Keller et al. demonstrated with MASCOT that this preprocessing can be performed efficiently using OT ~\cite{Keller2016}. Integrating QOT with SPDZ could further strengthen its security foundations.

In this work, we focus on two-party protocols and protocols that support dishonest majorities, such as Yao's protocol and SPDZ.
Both protocols are well-suited for integration with QOT, making them ideal candidates for implementation and further exploration.

\begin{table}[h!]
    \centering
    {
    \begin{tabular}{@{}ccccc@{}}
         \hline
         Protocol &Circuit & OT required (B)&Adversary &Parties\\
         \hline
         Yao  & B & 1 OT per input bit  & semi-honest & 2 \\
         \cite{Yao1986} & & (of one party) & & \\
         
         GMW  & B and A  & 1 OT per AND gate  & semi-honest & $\geq 2$ \\
         \cite{GMW87} &  & per couple of parties & & \\
         
         SPDZ  & B and A  & 1 OT per AND gate  & semi-honest  & $\geq 2$ \\
         \cite{spdz} &  & per couple of parties & and malicious & \\
         \hline
    \end{tabular}
    }
    \caption{Characteristics of the main MPC protocols secure in the dishonest majority setting. The circuit can be boolean (B) or arithmetic (A), the adversary can be semi-honest (passive) or malicious (active), as available in the MP-SPDZ~\cite{mp-spdz}. The comparison of OT required is in the case of boolean circuits since Yao based protocols work with only boolean circuits.}
    \label{tab:mpc_ot_integration}
\end{table}

\subsubsection{General Considerations}


For the implementation of the full system, it is essential to note that all protocols still require authenticated, and in some cases, private communication channels. These requirements are typically assumed within the security model. 
Consequently, even when replacing classical OT with QOT, additional mechanisms are necessary to ensure channel authenticity and, when required, privacy - thus eliminating the reliance on asymmetric cryptography.

To address this challenge, we integrate QKD in parallel with QOT, using the generated keys for robust authentication and, optionally, encryption of the classical communication channel.
Importantly, QKD can utilize the same physical layer as QOT, and both share similar components in the post-processing stage.
In this paper, we demonstrate how the quantum layer can be time-multiplexed to execute QOT and QKD in parallel, meeting the security requirements of MPC protocols at the communication layer.

\subsubsection{Sofware integration with MP-SPDZ}

To leverage state-of-the-art capabilities in MPC, we integrate QOT with MP-SPDZ, a widely used and versatile software framework for benchmarking MPC protocols ~\cite{mp-spdz}.
By replacing the framework's baseline OT implementation with our QOT stack, we enable seamless utilization of its full functionality.
The default $BaseOT$ protocol in MP-SPDZ is based on the work of \cite{lauter_simplest_2015}, which relies on the decisional Diffie-Hellman problem for security. By replacing this with our QOT implementation, every instance where the MPC protocol requires OT execution is handled via a QOT operation.

Since MPC applications often demand a large number of OTs, we employ OT extensions, as is standard in MPC implementations.
MP-SPDZ uses the SoftspokenOT extension ~\cite{Roy2022}, which  efficiently converts $s$ 1-out-of-2 OT of size $k$ into $k$ OT of size $s$.
In this scheme, $s$ depends on the security parameter, while $k$ is determined by the number of OTs for a particular computation.
Moreover, the string size $k$ can be fixed and extended to the desired size using a pseudo-random generator (PRG) during the extension process.
To run any MPC application within MP-SPDZ, we provide a fixed number of QOTs with a fixed string size. For instance, the standard configuration in MP-SPDZ requires 128 OTs of 128 bits, a requirement our QOT stack is designed to fulfill.


\section{Methods and Results}
This section summarizes the key methods and results from the practical implementation of the protocol and application described in the previous section. The physical layer features a polarization-entangled photon source operating in the telecom C-band, which distributes entangled photons to the two parties, Alice and Bob, participating in the oblivious transfer.
It is important to note that the physical layer and measurement protocol steps required up to the post-processing stage for the QOT implementation are identical to those used in the QKD-BBM92 protocol \cite{neumann2022continuous}. The two parties generate a raw key by measuring incoming photons in one of two unbiased, non-orthogonal bases, with the basis selection determined randomly by a 50/50 beam splitter. We use HV and DA bases, where H and V refer to horizontal and vertical polarization states, respectively, while D and A correspond to diagonal and anti-diagonal polarization states. Each receiver employs two polarization controller stages and two polarization beam splitters (PBS) to separate the incoming photons according to their polarization states $(\ket{H},\ket{V},\ket{+},\ket{-})$. When Alice and Bob measure in the same basis, they obtain fully correlated outcomes, enabling the transmission of one bit of information \cite{Bennett1992}.

We measure a coincidence rate of 28 kHz, considering all possible polarization basis combinations, by employing Super Conducting Nanowire Single-photon Detectors (SNSPDs).
The average single count rate per detector is 138.15 kHz. By calculating the ratio of coincidences to singles, we estimate the total system transmittance - accounting for all loss factors, from photon pair generation to detection efficiency - to be 2.5\%.
\begin{figure}[t]
\centering
\includegraphics[scale=0.65]{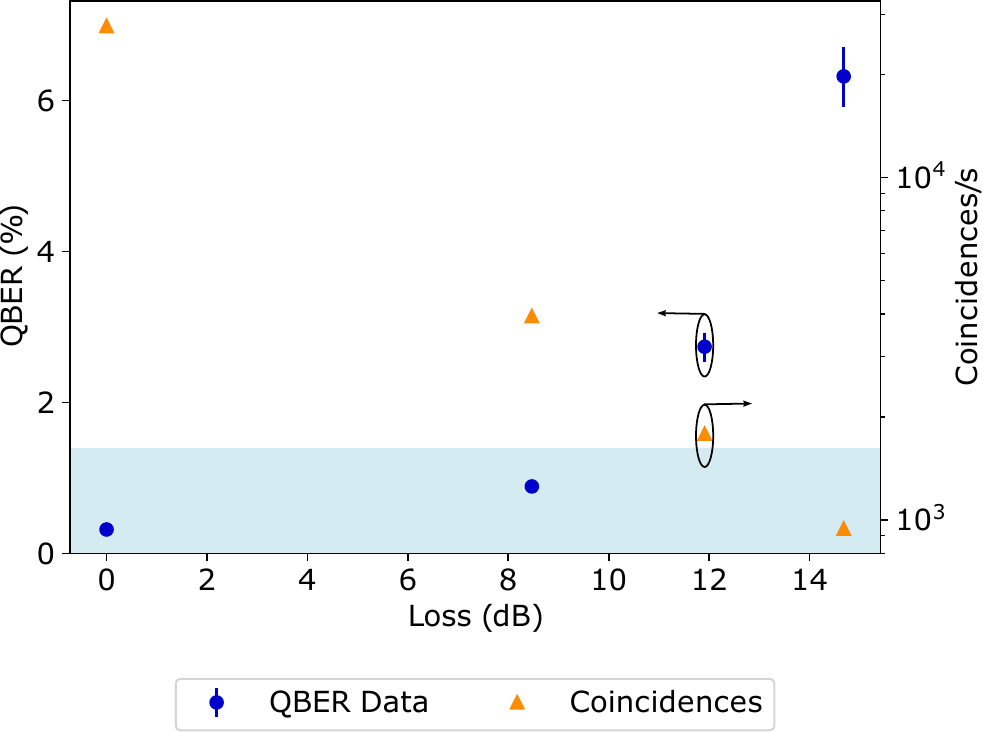}
\caption{QBER and coincidence rate per second of the entangled photon source as a function of optical loss (dB). Blue shadow ($p \leq p_{max}$) represents the area where OT keys can be generated. The error bars were calculated according with QBER boundaries defined in \cite{almeida_continuous_2016}.}
\label{fig:qber}
\end{figure} 
To quantify the degree of entanglement shared between Alice and Bob, the entanglement visibility is evaluated. This visibility measures how closely the generated quantum state matches to the ideal theoretical state \cite{Anwar2021}. Measurements of the entanglement visibility in back-to-back configuration (0 km fiber length) are provided in the Appendix \ref{sec:source}.
The average raw visibility in back-to-back is measured to be higher than $ 99\%$, indicating the high quality of the entangled photons used.

The QBER can be estimated from the measured entanglement visibility \emph{V} using the relation \cite{scheidl2009feasibility}
\begin{equation}
QBER = (1-V)/2.
\label{eq:QBER}
\end{equation}
The loss tolerance of the protocol is tested by distributing the entangled photons through fiber spools of varying lengths (0 km, 12.9 km, 17.2 km, 28.1 km) before splitting them (de-multiplexed in frequency) and sending them to the BBM92 receivers. 
Fig. \ref{fig:qber} shows the QBER and coincidences as a function of loss. As the loss increases, the signal-to-noise ratio (SNR) decreases, resulting in a higher QBER and a reduction in coincidences, which scales quadratically with the loss. Each data point is integrated over 1 minute. Polarization drifts in the fiber spools were found to be negligible, and the increased QBER can be attributed to the decrease in SNR. 
The QBER threshold to achieve a positive secure key rate for the used OT protocol is $p_{max} = 1.4\%$, see (\ref{eq:keyrate}).
As shown in Fig. \ref{fig:qber}, our system can generate secure OTs up to 8.47 dB loss corresponding to a fiber length of 25.8 km between parties.
In a back-to-back configuration, a long-term measurement is conducted over 13.2 hours, resulting in a QBER of $0.5 \%$ in the  HV basis and $1.2\%$ in the DA basis. Here, the QBER calculation of the source is performed only to estimate the loss and distance boundaries. For the remaining steps and the practical implementation of OT and MPC applications, the data was transmitted through the entire pipeline. These values are slightly higher than those reported in Fig. \ref{fig:qber}, likely due to mild instability during the long integration time, possibly caused by relaxation of the polarization controllers. Nevertheless, the average QBER was 0.75\%, which is well below the 1.4\% threshold needed to implement the protocol.



\bigskip



\subsection{Performance Analysis}\label{ssec:otperf}
\begin{figure}[t]
\centering
\includegraphics[scale=0.65]{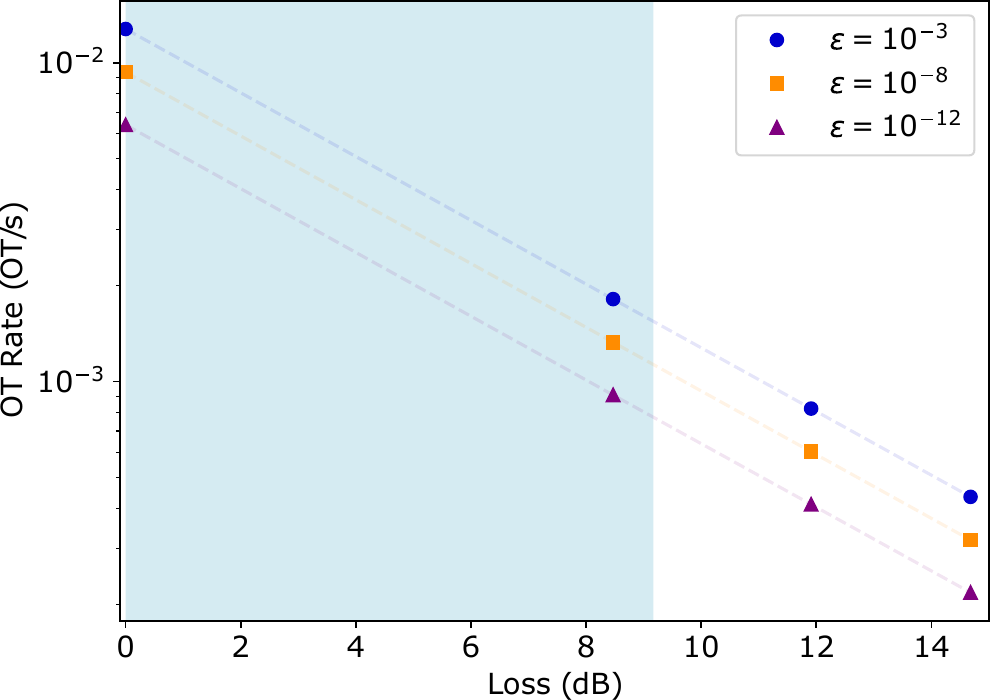}
\caption{OT rate in units of OTs per second as a function of optical fiber length for $\varepsilon = 10^{-3}, 10^{-8} \textnormal{ and } 10^{-12}$}
\label{fig:OTs}
\end{figure} 

In this subsection, we analyze the performance of the quantum OT protocol in terms of OT rate, quantum resource utilization, and loss tolerance for different security parameters $\varepsilon$, see Appendix \ref{sec:security}. Parameter $\varepsilon'$ depends solely on the security properties of the commitment scheme. In this analysis, we assume an adequately secure commitment scheme and focus specifically on the sum of the other two, $\varepsilon_{\textnormal{tot}} = \varepsilon +\varepsilon''$. Figure \ref{fig:OTs} shows the OT rate (OT per second) as a function of loss for various security parameters, under the assumption $p\leq p_{max}$. A larger $\varepsilon$ reflects a less stringent security condition, allowing higher OT rates. In a back-to-back configuration, the maximum OT rate of $1.3\times 10^{-2}$ OT/s was achieved for $\varepsilon=10^{-3}$, while the lowest rate of $6.4\times10^{-3}$ OT/s occurred for the highest security setting of $\varepsilon=10^{-12}$. With a reasonable security level of $\varepsilon=10^{-8}$, the protocol achieved an OT rate of $9.3\times 10{-3}$ OT/s in the back-to-back scenario and $1.8\times 10^{-3}$ OT/s at its maximum loss tolerance, corresponding to 25 km of fiber. This distance marks the maximum QBER threshold permitted by the protocol ($p_{max} = 1.4\%$). 

\begin{figure}[t]
\centering
\includegraphics[scale=0.65]{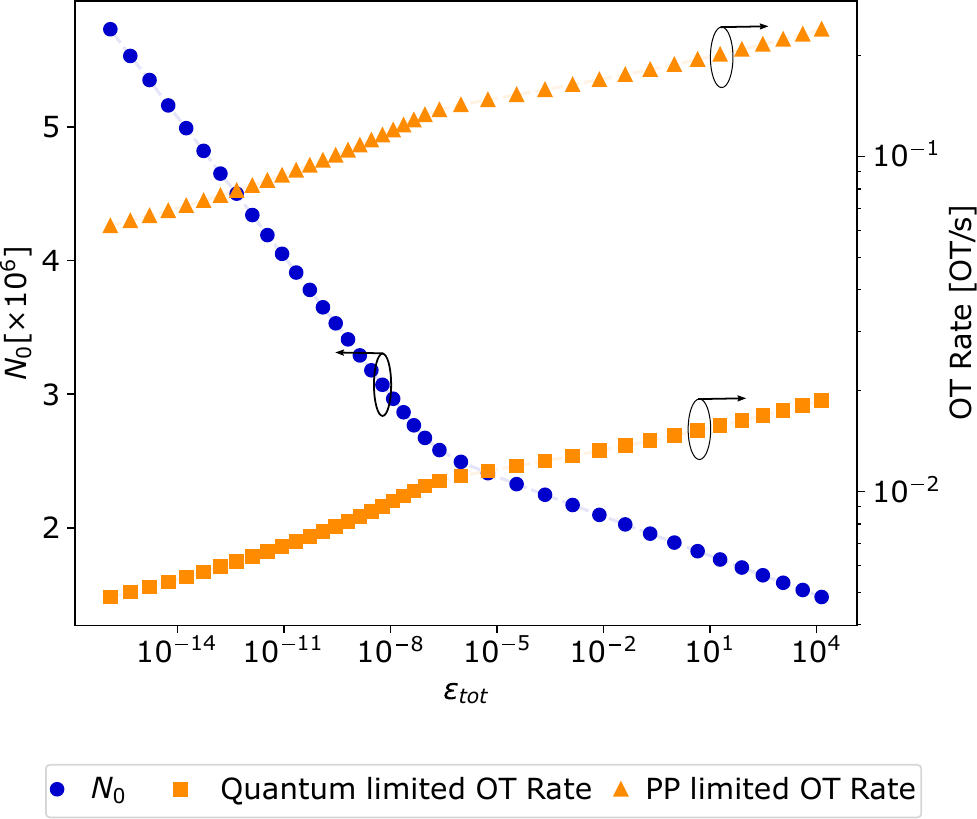}
\caption{$N_0$ and OT rate in OTs per second considering a pre-recorded data set ($\blacktriangle$), and a real-time implementation where the physical quantum system as the limiting factor ($\blacksquare$)}
\label{fig:N0_and_OTs}
\end{figure}

To obtain a $\varepsilon_{tot}=10^{-8}$ according to \cite{hernandez_asymmetric_2022, erven_experimental_2014, furrer_continuous-variable_2018}, we need an input block size of $3.2 \times 10^6$ events. With our optical setup providing a coincidence rate of 28.3 kHz this takes a recording time of 113 seconds. In an use case where real-time operation is required, the total performance is then $9.3\times10^{-3}$ OTs/s.
The most time consuming post-processing step is the bit commitment taking 38 seconds to compute one block. Considering a use case, where we can pre-record and buffer large amounts of measurement data, we can thus achieve a performance of 0.11 OTs/s. Figure \ref{fig:N0_and_OTs} illustrates the number of shared entanglement states $N_0$ as a function of security parameter, $\varepsilon_{tot}$, and the resulting OT rate for both real-time and pre-recorded offline processing configurations. The figure demonstrates that as the security parameter decreases (e.g., higher $\varepsilon_{tot}$) fewer shared entangled states are required, enabling higher OT rates. The performance gap between the pre-recorded and real-time scenarios is also evident. In the pre-recorded scenario, performance is limited by the computational capacity of the system handling the post-processing, while in the real-time scenario, the limitation arises from the constraints on the physical layer. This gap becomes increasingly pronounced for less stringent security parameters, where computational requirements are reduced. Conversely, for the highest security settings, the post-processing workload is significantly greater due to the increased volume of data, resulting in a heavier computational burden.

Considerable improvements at the physical layer can be implemented to reduce the gap and catch up with the post-processing latency. The broadband entangled photon source has been de-multiplexed with narrow-bandwidth filters (DWDM, 100GHz channel bandwidth). It is actually possible to use a spectral filter with a broader bandwidth (200GHz for example) without compromising the QBER, and therefore increasing the coincidence rate. Another improvement relies on increasing the power of the laser pumping the nonlinear crystal (see Fig. \ref{fig:source}). This typically comes with an increase of the QBER due to the onset of multi-pair events, so an optimal trade-off should be identified. From the detection side, improvements can be achieved by employing SNSPDs with better performances than the average $64 \%$ detection efficiency and 50~Hz dark count rate used in this work. Nowadays, high-efficient and low-noise SNSPDs are indeed commercially available. Moreover, ultrafast SNSPDs are also available which can reach a count rate up to 200MHz with an efficiency higher than $50 \%$, enabling thus high secret key rates at short distances \cite{stasi2024high}.   Lastly, to push the performances even further, a new source design leveraging a waveguide-based nonlinear crystal can be implemented in which the light-matter interaction is considerably enhanced compared to a source based on a bulk nonlinear crystal (this work) \cite{zhuang2024ultrabright}.

\subsection{Application Results}
After presenting the performance results of the physical layer and post-processing, this section evaluates the performance of the MPC application operating on top of the QOT protocol.
To implement the use case, we build on findings from \cite{mader_towards_2024} and adopt the approach recommended for efficient fingerprint matching using fixed-length templates, as proposed in \cite{engelsma_learning_2019}.
Matching is performed by computing the cosine similarity or the Euclidian distance between two vectors, both of which can be executed efficiently.
For pre-processing, we use the implementation from ~\cite{rohwedder_benchmarking_2023}, which recommends embedding lengths between 256 and 512.

To benchmark the complete software stack, we integrate QOT into MP-SPDZ and implement the MPC application for fingerprint matching.
Various configurations are tested and analyzed using the measurement data stored, as outlined below. 
In particular, we evaluate different problem sizes, defined by the number of fingerprints against which the candidate template is compared, as summarized in Table \ref{tab:FingerprintTable512}.
Here, $N=512$ represents the size of the vectors encoding the fixed-length fingerprint representation, where the unique and most relevant features of a fingerprint are captured in a 512-length vector \cite{rohwedder_benchmarking_2023}.
The similarity between this vector and each of the $M$ fingerprint vectors in the database is then measured in an oblivious manner within MPC protocol, preserving the privacy of both values.
Specifically, the comparison is based on the square of the Euclidean distance between the two vectors, requiring the computation of $N$ ~oblivious multiplications and $2N-1$ additions per database entry.
Notably, the execution of 128 QOTs takes $19$ minutes and $12$ seconds during the offline phase, based on the performance analysis in Section~\ref{ssec:otperf}.
In contrast, the ~execution of 128 classical OTs using the method from \cite{lauter_simplest_2015} takes approximately around 0.01 seconds.

\begin{table}[h!]
    \centering
    \begin{tabular}{@{}crcc@{}}
         \hline
         M  & \multicolumn{1}{c}{\#OT} & $t_{\textrm{off}}$ (h:m:s) & $t_{\textrm{on}}$ (s) \\
         \hline
         10 & 18343808 & 00:10:20 & 0.022 \\
         50 & 26165632 & 00:19:24 & 0.105 \\
         100 & 36318976 & 00:19:28 & 0.185 \\
         500 & 116291264 & 00:20:01 & 1.018 \\
         10000 & 2100776896 & 00:35:13 & 24.21\\
         100000 & 21012907136 & 03:01:23 & 242\\
         \hline
    \end{tabular}
    \caption{Private fingerprint matching with a feature vector length of N=512 is performed using the MASCOT protocol, where M is the size of the database. The $\#$OT represents the total number of OTs needed for the specific computation in the MPC function, which can be derived by extension of 128 OT.}
    \label{tab:FingerprintTable512}
\end{table}

Table \ref{tab:FingerprintTableProtocols} compares the execution of the program with different types of protocols. MASCOT (Malicious Arithmetic Secure Computation with Oblivious Transfer) refers to the original implementation in \cite{Keller2016}, 
Semi refers to MASCOT without all the steps required for malicious security, and Yao refers to Yao's garbled circuit protocol \cite{Yao1986}.
These protocols offer varying levels of security with trade-offs in efficiency, as shown in table \ref{tab:FingerprintTableProtocols} and explained previously in the protocol descriptions.
Note that Yao's protocol is a two-party protocol, and the total number of BaseOTs and OTs required is indicated in the table \ref{tab:FingerprintTableProtocols}. In contrast, MASCOT requires this number of BaseOTs and OTs to be executed between each pair of parties involved. For instance, in a two-party protocol between parties  $P_1$ and $P_2$, 128 OTs will be performed with $P_1$ as the sender and $P_2$ as the receiver, and another 128 OTs will be executed with $P_2$ as the sender and $P_1$ as the receiver. This does not increase the total execution time, as the two OT executions can be performed in parallel when the parties take on different roles.

\begin{table}[h!]
    \centering
    \begin{tabular}{@{}crcc@{}}
         \hline
         Protocol  & \multicolumn{1}{c}{\#OT} & $t_{\textrm{off}}$ (h:m:s) & $t_{on}$ (m:s:cs)  \\
         \hline
         MASCOT & 216483136 & 00:20:39 & 00:02:33    \\
         Semi & 33000320 & 00:19:24 & 00:01:47   \\
         Yao  & 16384000 & - & 21:11:37\\
         \hline
    \end{tabular}
    \caption{Comparison of the execution times for the fingerprint matching application using different protocols, with a database size of M=1000, vector length of N=512, and a BaseOT size of 128 bits.}
    \label{tab:FingerprintTableProtocols}
\end{table}

\section{Conclusions}

We have demonstrated a fully functional implementation of a MPC application supported by QOT. The QOT protocol is implemented on an entanglement-based physical-layer that uses polarization-encoded entangled states to share oblivious keys between two parties, and also enables QKD to provide authentication. Our implementation highlights versatility by combining QKD and QOT post-processing on the same physical layer. The authentication process between peer modules involves hashing messages into a crypto-context, verifying authenticity using authentication tags, and replenishing key material through a parallel QKD-pipeline when the pre-shared secret is exhausted, with QKD handling key post-processing and authentication to ensure continued operation.

The quantum OT protocol demonstrates statistical correctness and computational security for an honest receiver, as well as statistical security for an honest sender, under the assumption that the commitment scheme used is computationally hiding, statistically binding, and verifiably error-corrected. This provides a solid foundation for secure quantum communication in the context of oblivious transfer. For a security parameter of $\varepsilon= 10^{-8} $, oblivious keys are successfully generated over a maximum channel loss of 8.47 dB, corresponding to a distance of 12.9 ~km. In a back-to-back scenario with the same security parameter, we achieved a QOT rate of $9.3\times10^{-3}$ OTs/second, corresponding to approximately 1 minute and 48 seconds per OT. This rate is primarily constrained by the entanglement-source's ability to distribute quantum states efficiently. Moreover, generating 128 OTs cannot be achieved by simply multiplying this time by 128, as post-processing computations can be parallelized, thereby reducing their impact relative to the physical layer's operational time. 
For many applications, oblivious keys can be pre-distributed and stored in hardware until needed by the post-processing phase triggered by the MPC application. In such cases, using the same same security parameter as in the previous scenario, we achieve an OT rate of 0.11 OTs per second, corresponding to 9.1 seconds per OT. The generation of the 128 OTs can be assumed to incur no additional cost, as it is feasible to compute them in parallel. Compared with classical OT which is performed nearly instantaneously, the cost required for higher security, provided by quantum oblivious transfer, is 1 minute and 48 seconds in the first scenario and 9.1 seconds in the case of pre-recorded data. Notably, in the first scenario, this time could be further reduced through enhancements to the physical layer.

We present a practical demonstration of a fingerprint-matching use case, where a passenger boarding at a specific airport wishes to keep their identity and location private from the database-owning entities. The objective is to determine if a passenger's fingerprint matches any entry in a no-fly list maintained by Interpol and the United Nations. The fingerprint is secret-shared between both sites, ensuring ITS. The matching algorithm is computed using MPC, supported by a quantum oblivious transfer physical layer. The application uses 128 OTs obtained from post-processing and applies OT extensions to achieve the required number of OTs. The cost for achieving the highest level of security in this application is 20 minutes and 39 seconds.

\section*{Acknowledgements}
This work has received funding from the EU Horizon Europe Work Programme under project QSNP (no.
101114043). The authors acknowledge the employment of Superconducting Nanowire Single Photon Detector from Single Quantum via the EU Horizon Europe Qu-Test project (no. 101113901).
\appendix

\section{Quantum Physical layer}\label{sec:source}
\begin{figure}[t]
\centering
\includegraphics[scale=2]{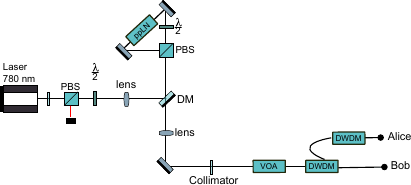}
\caption{Sketch of the type-0 broadband polarization entangled photon source. HWP: halfwave plate;
DM: dichroic mirror; PBS: polarizing beam splitter; PPLN: periodically poled
lithium niobate; VOA: Variable Optical attenuator DWDM:Dense Wavelength Division Multiplexing. }
\label{fig:source}
\end{figure} 

The entangled photon source used in this work is a type-0 Sagnac source, as illustrated in Fig. \ref{fig:source}. The source is pumped by a continuous-wave laser operating at 780 nm, with an average power of 40 mW. The pump beam is first passed through a polarization stage, where its polarization is adjusted to the diagonal state $\frac{1}{\sqrt{2]}}(|H\rangle+|V\rangle)$. The pump beam is then reflected by a dichroic mirror and directed into the Sagnac loop, which consists of a 25 mm-long periodically poled Lithium Niobate (ppLN) crystal for Spontaneous Parametric Down Conversion (SPDC) and a half-wave plate. 
A polarizing beam splitter (PBS) separates the pump beam into clockwise and counterclockwise directions. Both paths are focused into the SPDC crystal and recombine at the PBS. Due to the low efficiency of SPDC (about $10^{-9}$in this work), photon pairs are generated either in the clockwise or counterclockwise direction at any given time. By overlapping these paths and eliminating the which-path information, a maximally entangled Bell state is created.

\begin{figure}[t!]
\centering
\includegraphics[scale=0.75]{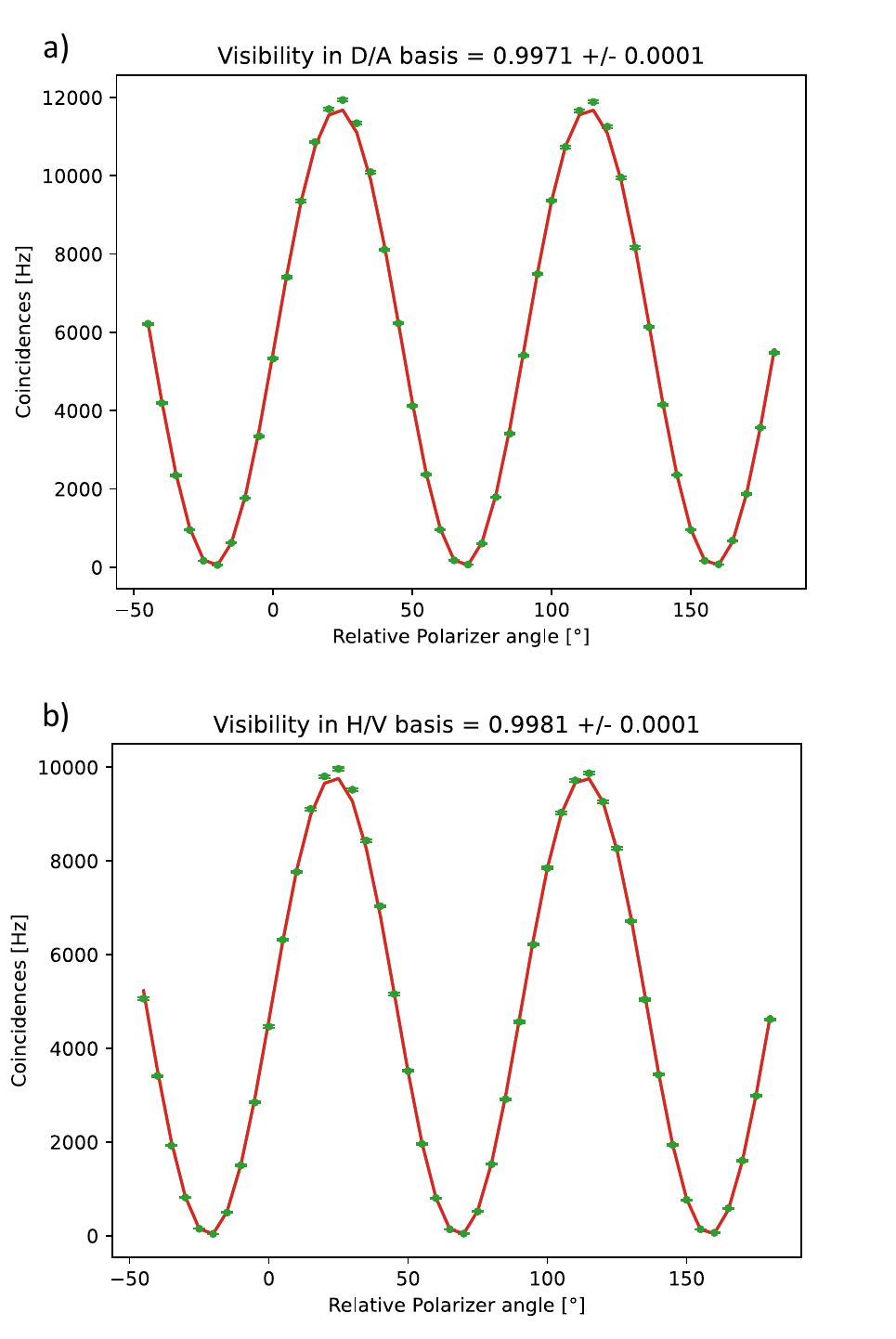}
\caption{Raw visibility in HV (a panel) and AD (b panel) basis. The green scatters are the experimental data, which are fitted
by the solid red line. Error bars are also shown
representing the standard error. }
\label{fig:vis}
\end{figure}

The SPDC spectrum has a Full Width Half Maximum (FWHM) of 85 nm, centered around the degenerate wavelength of approximately 1560 nm. By selecting frequency-correlated bands, the source can be spectrally demultiplexed into pairs of polarization-entangled photons.  
Wavelength demultiplexing was achieved using Dense Wavelength Division Multiplexing (DWDM) filters with central wavelengths of 1560.61 nm and 1558.17 nm. These filters have an insertion loss of 0.7 dB and an FWHM of 0.6 nm, respectively. The DWDM spectral filters correspond to channels 21 and 24 of the DWDM ITU grid. The DWDM filters correspond to channels 21 and 24 of the DWDM ITU grid. Further details about the entangled photon source can be found in \cite{trenti2024high}. As mentioned in the main text, various fiber lengths were included (before the DWDMs) to assess the maximum distance at which the protocol can be implemented. To evaluate the quality of the generated entangled photons, entanglement visibility curves were measured. The visibility parameter V is related to the Bell state fidelity by 
\begin{equation}
F > \frac{V(HV) + V(AD)}{2}
\end{equation}
Where V(HV) and V(AD) refer to the visibility measured in the HV and AD bases, respectively.
The entanglement visibility provides a lower bound for the fidelity, which indicates how closely the measured quantum state matches the theoretical state.
Two visibility curves were measured in both the HV and AD bases to assess the fidelity of the entangled photon source in a back-to-back configuration. These results are presented in Fig. \ref{fig:vis}. The measurements were performed using free-space entanglement analysis stages at both Alice and Bob's locations. Each analysis stage consists of two polarizers to set the desired polarization states for measurement.
The average visibility in the back-to-back configuration is $99.76\%$, indicating the high quality of the generated entangled photons.

For the oblivious transfer measurement, the BBM92 receiver setup is used, as shown in Fig. \ref{fig:smc}. Each receiver is equipped with four SNSPDs from Single Quantum. The detection efficiencies of the SNSPDs are $[62 \%,67 \%,64 \%,66 \%,70 \%,55 \%,69 \%,59 \%]$ with corresponding dark count rates (Hz) of $[<10, <100, <100, <50, <50, <50, <10, <50]$, respectively.

\section{Security of the OT Protocol}\label{sec:security}
The considered quantum OT protocol is a statistically correct, computationally secure for honest receiver, and statistically secure for honest sender assuming that: a)~the commitment scheme is both computationally hiding and statistically binding, and b)~the error correcting scheme is verifiable \cite{lemus_performance_2025}.

The statistical security of the protocol with security parameter $N_0$ is characterized by the values $\varepsilon,\varepsilon'$, corresponding to the correctness and security for an honest sender properties, respectively. More specifically, $\varepsilon$ denotes the statistical distance between the protocol's outputs and the ideal, correct, outputs when both parties are honest. Likewise, $\varepsilon'$, quantifies the distance between the protocol's final state and a state where the receiver is ignorant about at least one of the two output strings $m_0, m_1$ when the sender is honest. The expressions for $\varepsilon,\varepsilon'$ in terms of the protocol's parameters are given by

\begin{equation}
\begin{split}
\label{eq:idealsecurity}
\varepsilon &= 2^{-\frac{1}{2}(N_{\textnormal{raw}}-n)}   +2\varepsilon_{\textnormal{IR}} \\
\varepsilon' &= \sqrt{2} \left(e^{-\frac{1}{2} (1-\alpha)^{2} N_{\textnormal{test}} \delta^{2}_1} +  e^{-\frac{1}{2} N_{\textnormal{check}} \delta^{2}_1} \right)^{\frac{1}{2}} + e^{-D_{KL}(\frac{1}{2} - \delta_2|\frac{1}{2})(1-\alpha) N_0}   \\
&\quad \quad  + \frac{1}{2}\cdot 2^{n  - N_{\text{raw}} \left( \frac{1}{2} - \delta_2 - h\left( \frac{p_{\text{max}} + \delta_1}{\frac{1}{2} - \delta_2}\right) - f \cdot h(p_{\text{max}}+\delta_1) \right)},
\end{split}
\end{equation}

where $N_{\textnormal{test}} = \alpha N_{0}, N_{\textnormal{check}} = (\frac{1}{2}- \delta_2) \alpha N_{0}, N_{\textnormal{raw}} = (\frac{1}{2}- \delta_2) (1 - \alpha) N_{0}$, $\delta_1$ and $\delta_2$ are statistical tolerance parameters, $D_{KL}(\cdot | \cdot)$ denotes the binary relative entropy, and $\varepsilon_{\textnormal{IR}}$ is the probability that the error correction subroutine fails while undetected by the parties. \cite{hernandez_asymmetric_2022}.

\section{Security Proof for Commitment}\label{sec:appendix}

\subsection{Basic Commitment} \label{sec:bitcommitment}

Let $n$ denote the security parameter, and let $G$ represent a pseudo-random number generator (PRNG) that produces a string of length $3n$ from a seed of length $n$. The protocol's flow is shown in Table \ref{ourBSC}.
\begin{table}[t]
        \centering
        {
        \begin{tabular}{|l c l|}
            \hline
            Prover $\mathcal{P}$ & & Verifier $\mathcal{V}$ \\
            Commit to $\mathbf{b}=(b_1, b_2)$ & & \\
            \underline{Commit phase}: & & \\
            & & $\mathbf{r}_1 \xleftarrow{\$} \{ 0,1 \} ^{3n} \setminus\{\mathbf{0}, \mathbf{1}\}$ \\
            & $\xleftarrow{\hspace{20 pt}  \mathbf{r}_1 \hspace{20 pt}}$ & \\
            $\mathbf{r}_2 = \mathbf{r}_1 \gg 1 $ & & \\
            $\mathbf{x} \xleftarrow{\$} \{ 0,1 \} ^{n}$ & &\\
            $\mathbf{c} = G(\mathbf{x}) \oplus  b_1 \cdot \mathbf{r}_1 \oplus b_2 \cdot \mathbf{r}_2$ & &\\
            & $\xrightarrow{\hspace{20 pt} \mathbf{c} \hspace{20 pt}}$ & \\
            \underline{Reveal phase}: & & \\
            & $\xrightarrow{\hspace{20 pt} \mathbf{b}, \mathbf{x} \hspace{20 pt}}$ & \\
            & & Check that \\
            & & $\mathbf{c} = G(\mathbf{x}) \oplus b_1 \cdot \mathbf{r}_1 \oplus b_2 \cdot \mathbf{r}_2$ \\
            & & where $\mathbf{r}_2 = \mathbf{r}_1 \gg 1$ \\
        \hline
        \end{tabular}
        }
        \caption{Extension of Naor's bit CS from \cite{RLV24}.}
        \label{ourBSC}
    \end{table} 

The hiding property of this scheme is computationally secure, relying on the strength of the PRNG, while the binding property is statistically secure, offering $n-3$ bits of security. The same value of $\mathbf{r}_1$ can be reused across all protocol executions required to compute one OT, without compromising security, provided that $\mathbf{r}_1$ is refreshed for each new OT execution (i.e., it is not fixed). \newline
The proofs for the hiding and binding properties are presented in \cite{RLV24}, and are therefore omitted here.

\subsection{Using AES in CTR Mode}

\begin{figure}[h]
\centering
\includegraphics[trim=50 250 50 150,clip,width=1.3 \textwidth]{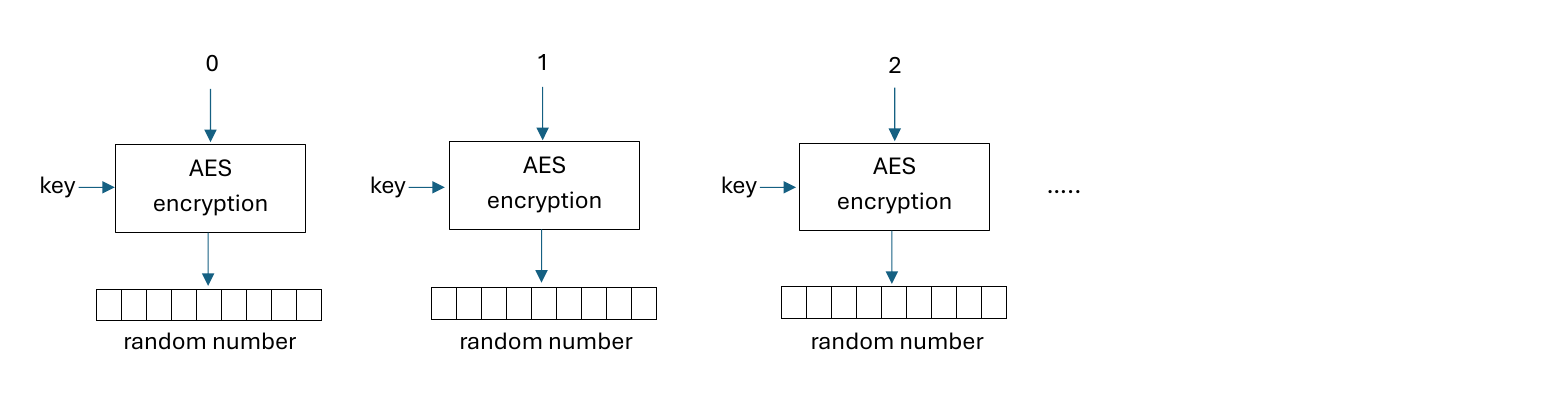}
\caption{AES in CTR mode as a PRNG.}
\label{AES_CTR_ENC}
\end{figure}

AES in counter mode (CTR) is used as a pseudo-random number generator. Specifically, the key serves as the seed, and the concatenation of ciphertexts forms the output.\\
Figure \ref{AES_CTR_ENC} illustrates the implementation of AES in CTR mode.
To generate the 768-bit string required by the commitment scheme (denoted as $G(\mathbf{x})$ in Fig. \ref{AES_CTR_ENC}), the verifier encrypts a counter, incrementing it by one after each encryption. Each encrypted value contributes to the string used in the commitment scheme. Specifically, six ciphertexts are concatenated to produced the required 768-bit string.

\subsection{Recap of the Entire Procedure in our Case}
The procedure is as follows:
\begin{itemize}
\item The verifier generates one random 768-bit string $\mathbf{r}_1$;
\item The verifier sends $\mathbf{r}_1$ to the prover;
\item For each pair of bits $(b_1, b_2)$ to be committed, the prover executes the following steps:
\begin{itemize}
\item[1.] Generates a random 256-bit AES key $\mathbf{x}$;
\item[2.] Encrypts the values 0, 1, 2, 3, 4 and 5 (using their 16 bytes representation);
\item[3.] Concatenates the results of the six encryptions to obtain the string $G(\mathbf{x})$;
\item[4.] Computes $G(\mathbf{x}) \oplus b_1 \cdot \mathbf{r}_1 \oplus b_2 \cdot \mathbf{r}_2$, where $\mathbf{r}_2 = \mathbf{r}_1 \gg 1$ and sends the result to the verifier;
\end{itemize}
\item To open a commitment, the verifier requests the AES key $\mathbf{x}$ and the two bits $(b_1, b_2)$. The verifier then computes the six encryptions using $\mathbf{x}$ to reconstruct $G(\mathbf{x})$, calculates $\mathbf{r}_2$, and verifies that the commitment matches $G(\mathbf{x}) \oplus b_1 \cdot \mathbf{r}_1 \oplus b_2 \cdot \mathbf{r}_2$.
\end{itemize}

\subsection{Exploiting a preprocessing phase to increase the efficiency of the bit commitment in the online phase}
The commitment in the online phase can be implemented with minimal additional effort by committing to random bits during a preprocessing phase. Specifically, suppose the prover has already committed to a bit $m$ in a preprocessing phase. During the online phase, it can commit to a new bit $b$ by publishing $m \oplus b$. \\
To open the commitment, the committer reveals the initial commitment to $m$, allowing the verifier to verify that commitment, disclose $m$, and compute $b$.

This method of committing to bits maintains the same security properties as the original bit commitment scheme used in the preprocessing phase: \begin{itemize}
    \item Hiding: the only way to deduce the committed bit $b$ in advance, is to learn either $b$ itself or the the bit it was XORed with (i.e., $m$, the bit committed to the preprocessing phase). Since the verifier has no means of guessing $b$, it would need to determine $m$, violating the hiding property of the preprocessing commitment.
    \item Binding: If the committer attempts to open the commitment to a different value (e.g., $b \oplus 1$), it must also open the initial commitment to $m$ as $m \oplus 1$, ensuring the XOR remains consistent. However, doing so would require breaking the binding property of the preprocessing phase. 
\end{itemize}
In this approach, a bit can be committed to in the online phase with the minimal cost of a single XOR operation. The same method can be extended to strings by committing to a random string during the preprocessing phase and XORing it the target string during the online phase).

\bibliographystyle{ieeetr}
\bibliography{apssamp,references-lore,references-mike, references-mariano} 

\end{document}